# Evaluating Wireless Proactive Routing Protocols under Mobility and Scalability Constraints

N. Javaid[1], R. D. Khan[2], M. Ilahi[1], L. Ali[1], Z. A. Khan[3], U. Qasim[4]

[1,2]COMSATS Institute of Information Technology, [1]Islamabad, [2]Wah Cant, Pakistan.
[3]Faculty of Engineering, Dalhousie University, Halifax, Canada.
[4]University of Alberta, Alberta, Canada.

**ABSTRACT**
Wireless Multi-hop Networks (WMhNs) provide users with the facility to communicate while moving with whatever the node speed, the node density and the number of traffic flows they want, without any unwanted delay and/or disruption. This paper contributes Linear Programming models (LP_models) for WMhNs. In WMhNs, different routing protocols are used to facilitate users demand(s). To practically examine the constraints of respective LP_models over different routing techniques, we select three proactive routing protocols; Destination Sequence Distance Vector (DSDV), Fish-eye State Routing (FSR) and Optimized Link State Routing (OLSR). These protocols are simulated in two important scenarios regarding to user demands; mobilities and different network flows. To evaluate the performance, we further relate the protocols strategy effects on respective constraints in selected network scenarios.

## I. INTRODUCTION

There are increasing interests in efficient routing in Wireless Multi-hop Networks (WMhNs). Due to self-organizing and self-configuring characteristics and absence of any infrastructural support made WMhNs an intense prospect in armed forces, disaster recovery areas, commerce, education and in many other applications. However, because of limited processing ability of nodes and constrained energy, the design of routing strategy in WMhNs is a challenging issue. The states of links are changing frequently broken due to wireless nature. Therefore, efficient routing is a big challenge in WMhNs.

Linear Programming is a mathematical technique which is used in computer modelling (simulation) to find the best possible solution in allocating limited resources to achieve maximum profit or minimum cost. However, it is applicable only where all relationships are linear (see linear relationship), and can accommodate only a limited class of cost functions.

WMhNs need efficient routing protocols which deal with dynamic topology producing less routing overhead. Many routing protocols have been proposed up till now. They are divided into two classes on the basis of their driven modes: table-driven protocols and on-demand driven protocols. In former category protocols, route discovery is originated from the traditional routing protocol, in which routing information between nodes is exchanged periodically, and each node maintains recent topological information. However, to provide paths quickly, high costs of routing overhead in terms of control packets are required to construct the routing tables with incorporated routing information. The main principle behind on-demand driven protocols is that the process of routing starts only when there are data to be sent. Therefore, routes are discovered only when the data request arrives.

In WMhNs, reactive protocols are responsible to find accurate routes and provide quick repair after detecting link breakages, whereas proactive protocols provide pre-computed routes without any delay of finding routes. This work is devoted to study the routing capabilities of three proactive protocols named as a Destination-Sequence Distance Vector (DSDV) [1] [2], Fish-eye State Routing (FSR) [3][4] and Optimized Link State Routing (OLSR) [5][6] in different network cases of WMhNs.

This paper contributes LP_models for performance parameters to evaluate selected routing protocols. We first list all the possible constraints of WMhNs for objective functions; throughput, cost of energy and cost of time. We evaluate selected protocols against these constraints. Moreover, we have also enhance default parameters of DSR and OLSR to achieve efficient performance. The contribution of this work includes: (i) construction of LP_models for WMhNs (ii) performance evaluation of selected protocols with respect to framework of network constraints, (iii) enhancement in DSR's and OLSR's default parameters, and (iv) analytical analysis of the mobility, traffic rates and scalability properties of selected routing protocols with 95% of confidence interval using NS-2.

## II.    RELATED WORK

In literature, we find different analysis on performance of routing protocols for different scenarios. A scalability analysis is presented in [7], which evaluates routing protocols with respect to different number of (CBR) resources. This analysis describes performance evaluation of AODV and DSR protocols influenced by the network size (up to 550 nodes), nodes' mobility and density. The authors in [8] evaluate the performance of DSR and AODV with varied number of sources (10 to 40 sources with different pause time). They demonstrate that even though DSR and AODV share a similar on-demand behavior, the differences in the protocol mechanics can lead to significant performance differentials. The problem from a different perspective is discussed in [8], using the simulation model with a dynamic network size and is examined practically for DSDV, AODV [9][10], DSR [11][12] and Temporally Ordered Routing Algorithm (TORA).

The authors in [13] examine the performance of proactive routing protocols. They set up a mathematical model to optimize proactive routing overhead without disturbing accuracy of routes. They present a generalized mathematical model for proactive routing protocol and specifically study the use of ACK mechanism. Finally they deduce that by optimizing the time interval of HELLO messages, proactive protocol will have less routing overhead and high delivery rate. Their evaluation based on mathematical model is generalized for proactive class, however, in our work, we specifically discuss the behavior of reactive (AODV, DSR, DYMO) along with proactive protocols (DSDV, FSR and OLSR).

In the study of [14], the authors compare the performance of AODV, OLSR and the Statistic-Based Routing (SBR) in terms of reliability and routing overhead of different traffic patterns using OPNET. They simulate OLSR and AODV as implemented in their respective RFCs except for the $V\_time$ (Valid time [11]) of Topology Control (TC) messages in OLSR which is decreased the value of `TC_INTERVAL` along with `HELLO_INTERVAL` to reduce the reaction time. In their work, they show that if a larger amount of overhead is taken into account, the protocols can achieve a slightly higher end-to-end reliability. Therefore they modify `HELLO_INTERVAL` and value of `ACTIVE_ROUTE_TIMEOUT` of AODV to compensate the frequent topology changes. In this paper, we evaluate OLSR and AODV along DSDV, DSR, DYMO and FSR. Moreover, like the work in [8], we also enhance OLSR.

In this paper, we formulate *LP_models*. These models list all possible constraints regarding selected objective functions; throughput, cost of energy and cost of time. Which protocol among selected protocols gives optimal solution in what scenario by satisfying *LP_model* constraints is discussed in detail by practically evaluating selected routing protocols in NS-2. The performance parameters for assessment of routing protocols are Throughput, Cost of Time and Cost of Energy. Different mobility rates, and varying densities scenarios are performed for evaluation in NS-2. For mobility evaluation, we have selected *2m/s* and *30m/s* with different pause times, whereas, for different network flows, varying numbers of nodes.

## III. PROBLEM FORMULATION USING LINEAR PROGRAMMING

We formulate *LP_models* for performance metrics; throughput, energy cost and time cost for WMhNs. These models are discussed below in detail:

## IV. LP_MODEL FOR MAXIMIZING THROUGHPUT (max $T\_avg$)

A protocol is aimed to provide efficient data delivery by end-to-end path calculations. These parameters along with their effects on objective function Max $T\_avg$ are discussed below:

- $dr$ Denotes an individual data request in a set of all data requests ($DR$); ($dr \in DR$)
- $\tau$ and $T$ specify unit time and simulation time, respectively (whereas $(\tau \in T)$.
- $Rec_{dr}$ is the number of successfully received data packet(s). Only $Rec_{dr}$ is considered for throughput measurements.
- Routing protocols are supposed to provide accurate routes for each $dr$. $p_{nr}$ represents the probability of no route for $dr$ during route discovery/route calculation process.
- $\tau^{(rp)}_{(out\_b)}$ is the buffer time out for a routing protocol and it can be less than or equal to the maximum time allowed for a data request to be buffered; $\tau_{max\_allowed}$.
- let $\alpha$ is rate parameter, so, data request arrival(s) and successfully received data packets rates and are represented by $\alpha_{tra}$ and $\alpha_{rec}$ respectively. Generally, $\alpha_{tra}$ is the data request transmission rate by the source node, while $\alpha_{rec}$ is the rate of received data packets rates at destination node.
- $\beta_{avail}$ is the available bandwidth of the channel during $\tau$.
- In wireless communication, the links between the nodes are frequently connected and disconnected. In $lb \in LB$, the object $lb$ represents the link breakage rate at any instant time $\tau$, and $LB$ symbolizes the whole link breakage rate during all the network connectivity period ($T$).
- In wireless communication, a network graph $G(V, E)$; here $V$ is the vertices and $E$ represents the edges or links between the nodes in the connectivity of the network graph. Any two nodes which are within the maximum allowable transmission rang $R^{max}_{(i,j)}$ i.e., in other words, it is necessary the difference of distance between the upstream and downstream links is less than or equal to $R^{max}_{(i,j)}$ for a node pair in a connected network. $LC_{max}$ is the maximum number of link change values during the connectivity period of a network.
- $\alpha^{(pro)}_{lr}$ Is the link repair response rate produced by a routing protocol correspond to each $r$.
- $rn$ Represents a node in a route among a set of all active routes; $RN$.
- $p^{(pro)}_{up\_rtable}$ Denotes the probability of updates routing table of proactive protocols.

Here, we are considering only received packets for throughput measurements. Thus the objective function $max\, T_{avg}$, is expressed as:

$$\text{Max}\, T_{avg} = \frac{\sum_{r \in R}(1 - p_{nr}) Rec_r}{\sum_{\tau \in T} \tau} \left(\frac{B}{s}\right) r \in R \qquad (1)$$

**Subject to**

$$\forall lb \in LB \qquad \sum_{r \in R} \alpha_{lr}^{(pro)} \leq LC_{max} \qquad (1.a)$$

$$\forall lb \in LB \qquad dist_j - dist_i \in R_{(i,j)}^{max} \qquad \forall (i,j) \in E \qquad (1.b)$$

$$p_{up_{rtable}}^{pro} \leq 1 \qquad \forall r \in R \qquad (1.c)$$

## V. LP_MODEL FOR ROUTING DELAY CT

Routing delay; $CT$ is the time required by a routing protocol for processing incoming $dr$ followed by RD and RM time costs.

• $\tau_{cri}$ Stipulate the critical delay value, which means that remaining is not enough to further transmission for $r$. Such a situation arises in case of delay in the route discovery in dense network, high data rates and high mobilities due to extensive link breakage. All these situations can result buffer_time_out.

• For minimizing delay the interval between periodic updates ($\tau_{per\_up}$) must be less than critical time $\tau_{cri}$. Moreover, for convergence purpose, $LSM$ an interval $\tau_{ls\_mon}$ as well as the time required to perform trigger updates ($\tau_{tr\_up}$) must be less than $\tau_{cri}$.

Let $CT$ be the required minimizing objective function used to express routing delay generated by reactive routing protocols, we write this as:

$$\text{Min } CT \qquad (2)$$

**Subject to**

$$\sum_{\forall l \in NB} \tau_{LSM}^{rp} < \tau_{cri} \qquad (2.a)$$

$$\sum_{\forall r \in RT} \tau_{RU_{per}}^{rp} < \tau_{cri} \qquad (2.a)$$

$$\sum_{\forall l \in r} \tau_{RU\_tri}^{rp} < \tau_{cri} \qquad (2.a)$$

## VI. LP_MODELING FOR ROUTING OVERHEAD CE

The parameters along with their effects on objective function $(min\ CE)$ are discussed below:

• Routing overhead; $CE$ represents the number of routing packets produced by a routing protocol and it depends upon the nature and operations of protocols.

• $\beta_{cri}$ stipulates the critical bandwidth which restricts further transmission for $dr$ data. Such a situation arises in case of high data rates and high mobilities.

• Proactive protocols can perform three types of maintenance operations for calculations and maintenance of routes; Link State Monitoring $(LSM)$, Route Updates periodically $(RU\_per)$ and Route Updates triggered $(RU\_tri)$. In our model, we take number of routing packets in term of energy cost, $CE$. Depending upon the combination of selected maintenance operations of a protocol,

$CE$ differs each other.

- $l$ Represents link(s) in $r$. Proactive protocols maintain link information by checking the connectivity of links periodically, this process is known as LSM. $EC$ for LSM represented by $EC_{LSM}$.

- $p_{lb}$ Denotes breakage probability of $l$.

Let $min\ CE$ is the minimizing objective function used to express routing overhead generated by reactive and proactive routing protocols. We can write this as:

$$\text{Min } CE \tag{3}$$

**Subject to**

$$\forall r \in RT \quad \forall r \in R; CE_{ls\_mon} < \beta_{cri} \tag{3.a}$$

$$\forall l \in r \quad CE_{per\_up} < \beta_{cri} \tag{3.b}$$

$$\forall nr \in NR\ (p_{lb})CE_{tri_{up}} < \beta_{cri} \tag{3.c}$$

## VII. PROACTIVE PROTOCOLS WITH THEIR BASIC OPERATIONS

As we discuss above that for analysing the effect of network constraints, we have selected DSDV, FSR and OLSR. Respective maintenance operations of these protocols are given below:

## VIII. DSDV

DSDV protocol performs three types of maintenance operations; $LSM$, $RU\_per$ and $RU\_tri$ as mentioned in [15]. Whereas, this protocol sends routing messages for $RU\_tri$ and $RU\_per$, because of link sensing from MAC layer. So, $CE$ of DSDV depends on the interval of $RU\_per$ and $RU\_tri$. Moreover, DSDV uses flooding mechanism to disseminate routing information. Let $CE_{RU\_per}^{DSDV}$ and $CE_{RU\_tri}^{DSDV}$ represents $CE$ of periodic and trigger updates of DSDV, as shown in Fig. 1(a) block A,B, and we can write the total $CE$ as:

$$CE_{total}^{DSDV} = CE_{RU\_per}^{DSDV} + CE_{RU\_tri}^{DSDV} \tag{4}$$

$$CE_{RU\_per}^{DSDV} = \left(\frac{\tau_{NL}}{\tau_{RU_{per}}}\right)\sum_{i=1}^{N} i \tag{4.a}$$

$$CE_{RU\_tri}^{DSDV} = \int_{\tau_{NS}}^{\tau_{NL}} sgn|S_{lb}^{AR}|\sum_{i=1}^{N} i \tag{4.b}$$

where, generation of $RU_{tri}$ depends on status of $lb$ among $AR$.

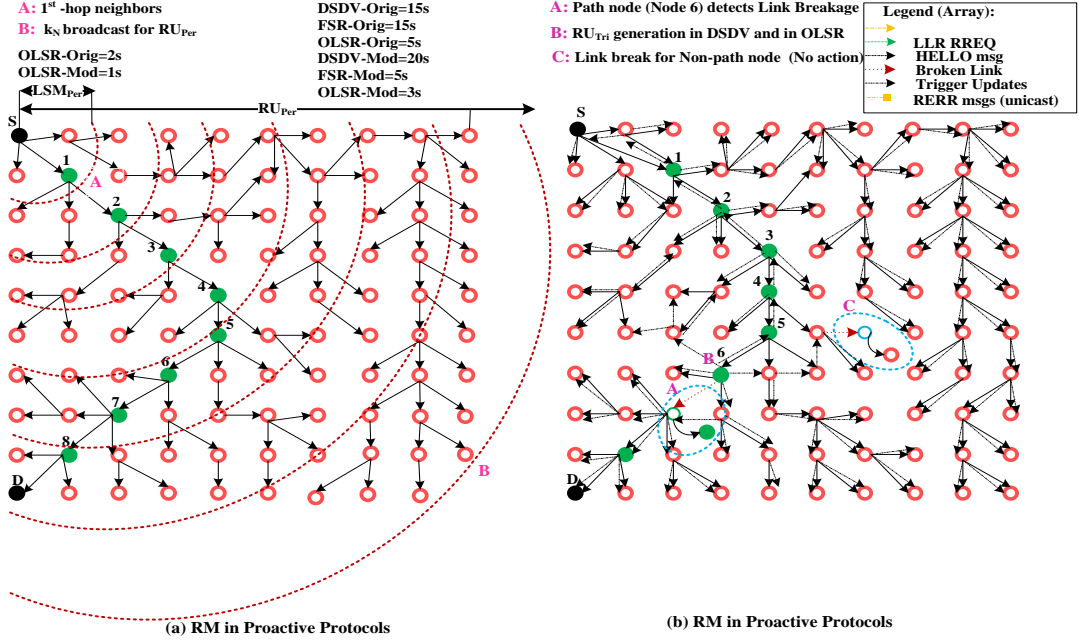

(a) RM in Proactive Protocols  (b) RM in Proactive Protocols

Fig. 1. Routing Operations in proactive Protocols

========================================================
**Algorithm. 1: Routing Operations in DSDV**
========================================================

**Preconditions:** Update_Interval = Periodic ($\tau_{per}$), Trigger ($\tau_{tri}$)

**Begin**

**If** Update_Interval = $\tau_{per}$ **then**

**Forall** $n \in N$ **do**

$S_c^{RT} \leftarrow$ Size changes in Routing Table (RT)

$S_b^{NPDU} \leftarrow$ Size of broadcasting NPDU

**If** ($S_c^{RT} > S_b^{NPDU}$) **then**

 Insert RT_INFO in NPDU
Transmit NPDU

$S_{(r-info)}^{RT} \leftarrow S_c^{RT} - S_b^{NPDU}$

**While** $S_{(r-info)}^{RT}$ **do**

 Insert RT_INFO in NPDU
Transmit NPDU

$S_{(r\_info)}^{RT} \leftarrow S_c^{RT} - S_b^{NPDU}$

**End while**
**else**

 Insert RT_INFO in NPDU
Transmit single NPDU

**End if**
**End for**
**else**
**Forall** $n \in N$ **do**

**If** $(c^{AR})$ **then**

Update_Interval ← $\tau_{tri}$

Sequence_Number ← ∞

Broadcast NPDU

**else**

Perform only periodic updates
**End if**
**End for**
**End if**
=========================================================
**Algorithm. 2. Maintenance Operation in FSR**
=========================================================
**Preconditions:** Update_Intervals={periodic for IntraScope($\tau_{IAS}$), periodic for InterScope ($\tau_{IES}$)}

**Begin**

**If** Update_Interval=$\tau_{IES}$ **then**

**Forall** $n \in N$ **do**

Set TTL ← 2

Broadcast Link State INFO
**Else**
Update_Interval=$\tau_{IES}$

**Forall** $n \in N$ **do**

Set TTL ← 255

Broadcast Link State INFO
**End for**
**End if**
=========================================================
**Algorithm. 3. Maintenance Operations in OLSR**
=========================================================
**Preconditions:** Update_Interval = { per_HELLO, TC Default (TC_def), TC Trigger (TC_trig)}

**Begin**

**If** Update_Interval=per_HELLO **then**

**Forall** $n \in N$ **do**

Set TTL ← 1 in HELLO message

Broadcast HELLO messages
**End for**

**Forall** $n \in N$ **do**

Calculate MPRs
**End for**

**Elseif** Change in Status of MPRs $(c^{MPR})$ occurs **then**

Update_Interval ← TC_tri

**Forall** $n \in N$ **then**

Broadcast TC messages
**End for**
**Else**

**Forall** $n_{MPR} \in N_{MPR}$ **do**
After expiring TC_def Interval
Update_Interval ← TC_def
Broadcast TC messages
**End for**
**End if**
======================================================

### IX. FSR

To avoid routing overhead, FSR only uses periodic maintenance operations; $LSM$ and $RU\_per$. For $LSM$ and $RU\_per$, MAC layer notification and Scope Routing *(SR)* are performed, respectively. In *SR*, diameter of whole network is divided into scopes and information is exchanged between scopes using graded-frequency technique. Two scopes; Inter-Scope and Intra-Scope are defined for FSR in [2] and $CE$ for these scopes is given below:

$$CE_{total}^{FSR} = CE_{RU\_per}^{IAS} + CE_{RU\_per}^{IES} \qquad (5)$$

$$CE_{RU\_per}^{IAS} = \left(\frac{\tau_{NL}}{\tau_{IAS}}\right) \sum_{i=1}^{N} \sum_{i=1}^{N_{IAS}} i \qquad (5.a)$$

$$CE_{RU\_per}^{IES} = \left(\frac{\tau_{NL}}{\tau_{IES}}\right) \sum_{i=1}^{N} \sum_{i=1}^{N_{IES}} i \qquad (5.b)$$

Here, $\tau_{IAS}$ and $\tau_{IES}$ are IntraScope_Interval and InterScope_Interval, respectively (Table. 1). Whereas, $N_{IAS}$ and $N_{IES}$ represent total number of nodes in IntraScope (IAS) and InterScope (IES).

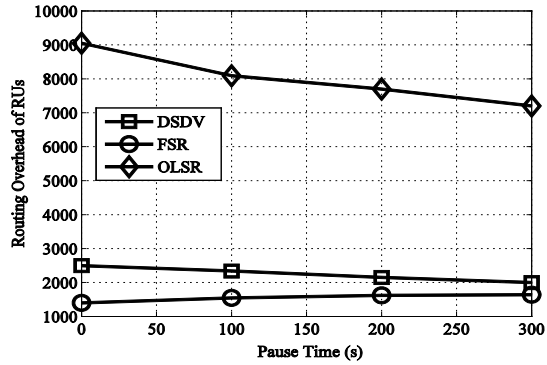

Fig. 2. Analytical Simulations of Mobility in Proactive Protocols

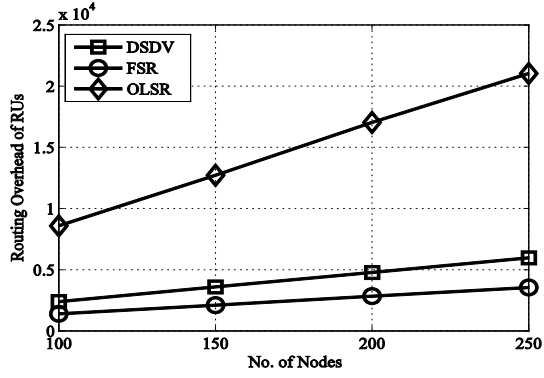

Fig. 3. Analytical Simulations of Scalability in Proactive Protocols

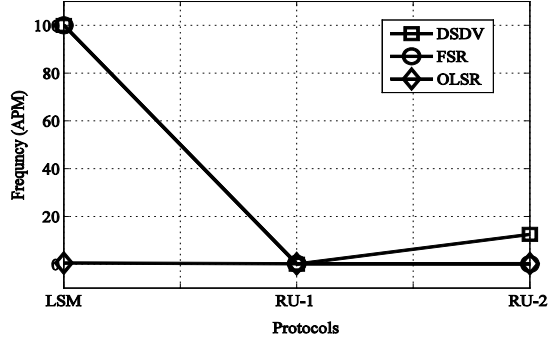

Fig. 4. Analytical Simulations of Update Frequency in Proactive Protocols

## X. OLSR

In OLSR, $LSM$ and $RU\_tri$ are used to get information for links and routes. $LSM$ is performed by generating HELLO messages on routing layer after $HELLO\_INTERVAL$ ($LSM$ in Fig. 7 and Table. 1). Whereas, $RU_{tri}$ is broadcasted through TC messages. The interval between successive $RU_{tri}$ depends on stability of MPRs. This stability is periodically confirmed through HELLO messages. On the other hand, to calculate topology information, TC messages are broadcasted. The broadcasting period of TC message depends on status of MPRs after $TC\_INTERVAL$ (default value as mentioned in Table 1) if MPRs are stable, while these messages are triggered and are transmitted to whole network in case of unstable MPRs, as portrayed in Fig. 1(b), when node 6 detects link breakage then OLSR generates $RU_{tri}$. The $CE$ of OLSR is given below:

$$CE_{total}^{OLSR} = CE_{LSM}^{OLSR} + CE_{RU\_tri}^{OLSR} \qquad (6)$$

$$EC_{LSM}^{OLSR} = \left(\frac{\tau_{NL}}{\tau_{HELLO}}\right)\sum_{i=1}^{N} nb_i \qquad (6.a)$$

$$EC_{LSM}^{OLSR} = \begin{cases} \int_{\tau_{NS}}^{\tau_{NL}} \sum_{i=1}^{MPRs} i & \text{If MPRs are stable} \\ \int_{\tau_{NS}}^{\tau_{NL}} \sum_{i=1}^{N} i & \text{Otherwise} \end{cases} \qquad (6.b)$$

**Table 1: Predefined Parameters Values**

| Parameters | Used by (Protocol(s)) | VALUES |
|---|---|---|
| $RU_{per}$ Interval | DSDV | 15s |
| $LSM$ of MAC Interval | DSDV, FSR | 0.1s to 0.8s |
| HELLO_INTERVAL | OLSR | 2s |
| TC_INTERVAL (default) | OLSR | 5s |
| TTL value for IntraScope | FSR | 2-hops |
| IntraScope_Interval | FSR | 5s |
| TTL value for InterScope | FSR | 255-hops |
| InterScope_Interval | FSR | 15 |

## XI. SIMULATIONS AND DISCUSSIONS

To evaluate chosen protocols, we take different mobilities, scalabilities. We selected three performance metrics; throughput, $CT$ and $CE$. We analytically simulate $CT$ in terms of routing overhead and $CE$ in terms of frequency of topological exchange periods in Figs. 5-7. The performance metrics are measured through simulations in NS-2. For simulation setup, Random Way-Point is used as mobility models. The area specified is $1000m \times 1000m$ field presenting a square space to allow mobile nodes to move inside. All of the nodes are provided with wireless links of a bandwidth of $2Mbps$ to transmit on. Simulations are run for $900s$ each. For evaluating mobilities effects, we vary pause time from $0s$ to $900s$ for $50$ nodes with speed $30m/s$. For evaluating different network flows with $15m/s$ speed and fixed pause time of $2s$, we vary nodes from $10$ nodes to $100$ nodes.

## XII. THROUGHPUT

Among proactive protocols, DSDV attains the highest throughput and shows efficient behavior in all pause times for, as shown in Fig. 8. The reason for this good throughput is to use of route settling time; when the first data packet arrives, it is kept until the best route is found for a particular destination, thus overall satisfied constraints. Secondly, a decision may delay to advertise the routes which are about to change soon, thus damping fluctuations of the route tables. The rebroadcasts of the routes with the same sequence number are minimized by delaying the advertisement of unstabilized routes. This enhances the accuracy of valid routes and thus satisfies constraint in eq. 1.f. resulting in the increased throughput of DSDV in all types of mobility rates, moreover, the updates are transmitting through NPDU's in small scalabilities. Whereas, due to low convergence of OLSR in high mobility Fig. 12, decreases overall throughput. The reason for this gradual decrease with increasing mobility is the unavailability of valid routes due to its proactive nature. In static situation as well as low speed, in Fig. 8, throughput is better as compared to moderate and relatively high mobility due to availability of stable entries for MPRs (eq. 6.b).

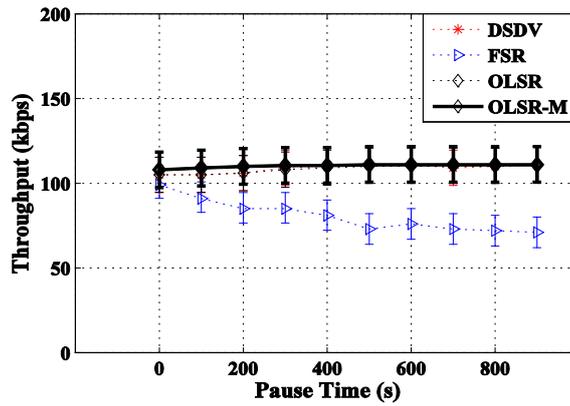
Fig. 5. Throughput vs Pause Time of Proactive protocols

Thus, in moderate and no mobilities OLSR performs well (Pause time more than $400s$ represents moderate mobilities, while pause time $900s$ means static mobile, because total simulation time is $900s$. Moreover, FSR does not trigger any control messages unlike DSDV and OLSR when links breaks, as depicted in Fig. 6. Therefore, it is not as efficient as DSDV and OLSR.

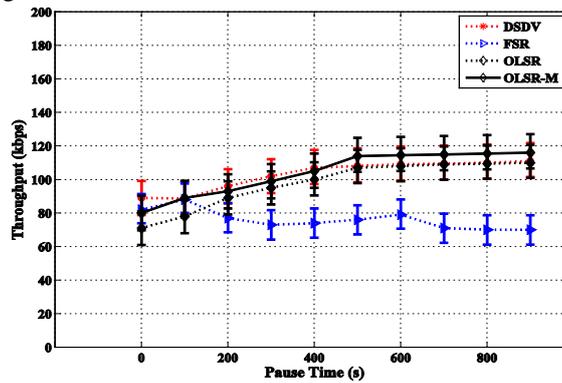
**Fig. 6.** Throughput vs Pause Time of Proactive protocols

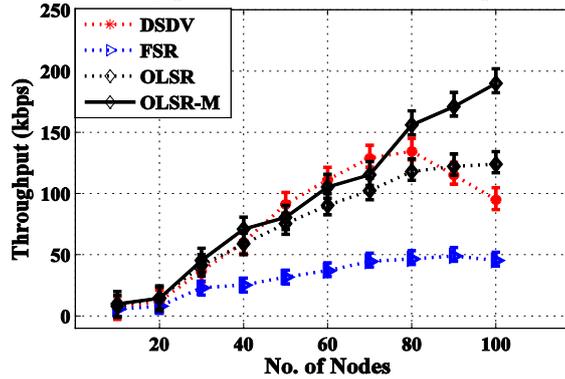
Fig. 7. Throughput vs Scalability of Proactive protocols

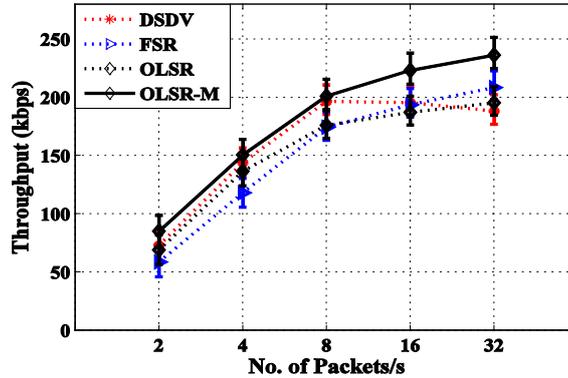

Fig. 8. Throughput vs Transmission Rate of Proactive protocols

FSR shows appreciable performance for varying traffic rates and OLSR is well scalable among proactive protocols. In medium and high traffic loads, FSR's performance is depicted in Fig. 10 and Fig. 11. This is due to introduction of new technique of multi-level Fish-eye Scope (FS), that reduces routing overhead and works better when available bandwidth is low, thus increasing throughput in case of increased data traffic loads and reduces routing update overhead. Although, DSDV uses NPDUs to reduce routing transparency but $RU_{tri}$ causes routing overhead and degrades performance. OLSR uses MPRs for reduction of overhead but computation of these MPRs takes more bandwidth. Therefore its throughput is less than FSR. Further optimization helps FSR to only broadcast topology messages to neighbors in order to reduce flooding overhead. If FSR would have taken MAC layer feedback in case of link brakes then there might be exchange of messages to update neighbors, consuming bandwidth and lowering throughput. This faster discovery results in a better performance during high traffic loads. Simulation results of OLSR in Fig. 10 comparative to Fig. 11 show that it is scalable but less converged protocol for high traffic rates. This protocol is well suited for large and dense mobile networks, as it selects optimal routes (in terms of number of hops) and achieves more optimizations using MPRs. OLSR-M due to exchanging information of neighbors and with topology through frequent exchange results more throughput, as shown in Figs. 8-11.

## XIII. COST OF TIME

In all proactive protocols, $CT$ value is directly proportional to speed and mobility, as depicted in Fig. 12 and Fig. 13. DSDV possesses the highest delay cost among proactive in moderate and no mobility situations, as well as in all cases its E2ED is higher than OLSR. Because in DSDV, a data packet is kept for the duration between arrival of the first packet and selection of the best route for a particular destination. This selection creates delay in advertising routes which are about to change soon, thus causing damping fluctuations of the route tables. Furthermore, advertisement of the routes which are not stabilized yet is delayed in order to reduce the number of rebroadcasts of possible route entries that normally arrive with the same sequence number. FSR at higher mobilities produces the highest $CT$ value among proactive protocols. Due to graded-frequency mechanism when mobility increases, routes to remote destinations become less accurate. However, when a packet approaches its destination, it finds increasingly accurate routing instructions as it enters sectors with a higher refresh rate. At moderate and no mobilities at all speeds, the value of end to end delay is the same as well as this delay is less than other proactive protocol due to SR.

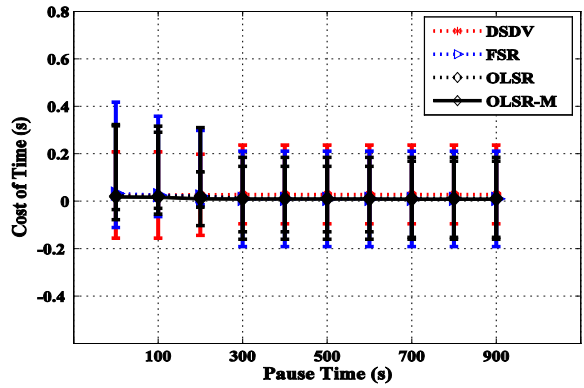
Fig. 9. Time Cost vs Pause Time of Proactive protocols

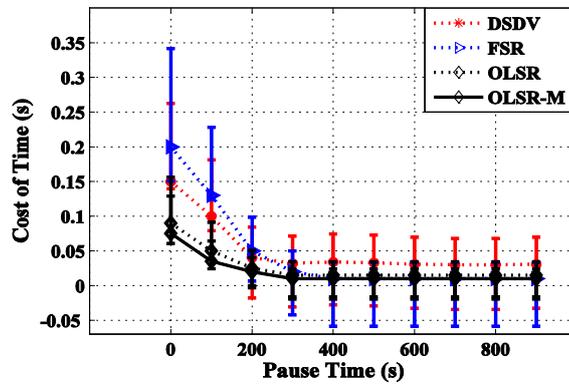
Fig. 10. Time Cost vs Pause Time of Proactive protocols

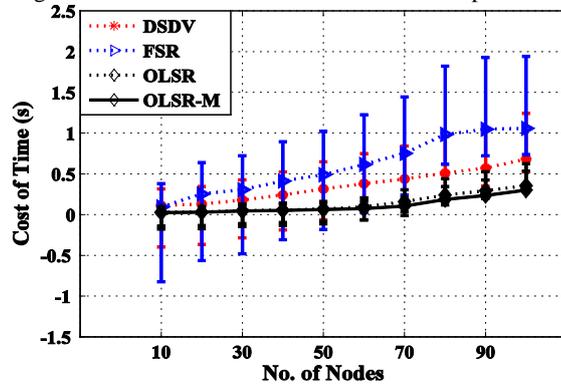
Fig. 11. Time Cost vs Scalability of Proactive protocols

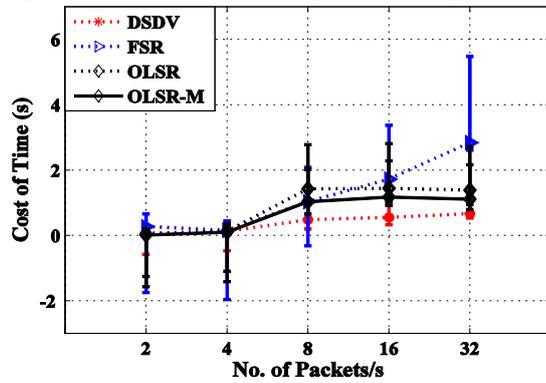
Fig. 12. Time Cost vs Transmission Rate of Proactive protocols

FSR overall suffers higher delay in scalabilities due to retain route entries for each destination, this protocol maintains low single packet latency when population is small as shown in Fig. 14 and Fig. 15. The graded-frequency mechanism is used to find destination to keep routing overhead low. FSR exchanges updates more frequently to the near destinations. Thus, in higher data rates or more scalabilities this protocol attains more $CT$ value. The reason for delay in DSDV is that it waits to transmit a data packet for an interval between arrival of first route and the best route. This selection creates delay in advertising routes which are about to change soon. A node uses new entry for subsequent forwarding decisions and route settling time is used to decide how long to wait before advertising it. This strategy helps to compute accurate route but produces more delay. Small values of $CT$ for OLSR are seen among proactive protocols in all scalabilities, because, MPRs provides efficient flooding control mechanism; instead of broadcasting, control packets are exchanged with neighbors only. In OLSR-M, routing latency is further decreased as compared to OLSR due decreasing $RU_{Tri}$ and $LSM_{Per}$ intervals (In Fig. 1(a) and Figs. 12-15).

### XIV. COST OF ENERGY

Figures 16 and 17 show that OLSR and OLSR-M due to computation of MPRs through TC and HELLO messages results in the highes generation rate of routing packets. The lowest $CE$ value is produced by DSDV, because, incremental and periodic updates through NPDUs reduce the routing overhead. Moreover, FSR has lower routing overhead than OLSR because it prefers periodic updates instead of event driven exchanges of the topology map which greatly helps in reducing the control message overhead during high mobility rates. Also, in FSR link state packets are not flooded. Instead, nodes maintain a link state table based on the up-to-date information received from neighbor nodes and are periodically exchange it with their local neighbors only (no flooding), as shown in Fig. 16 and 17.

As depicted in Fig. 18 and Fig. 19 in all scalabilities and traffic loads, OLSR and OLSR-M generate the highest NRL among proactive protocols (in eq. 3 constraints 3.a,c). It happens due to MPR mechanism that controls the dissemination of control packets in the whole network. But calculation of these MPRs through TC messages and HELLO messages increase the routing load. DSDV and FSR sustain low overhead in all network loads and in low and medium scalabilities. As, DSDV upholds routing table with separate route entry for new destination, while a node does not use the new entry for the same destination in making subsequent forwarding decisions. Moreover, NPDUs are arranged to disseminate incremental updates for maintaining low routing overhead.

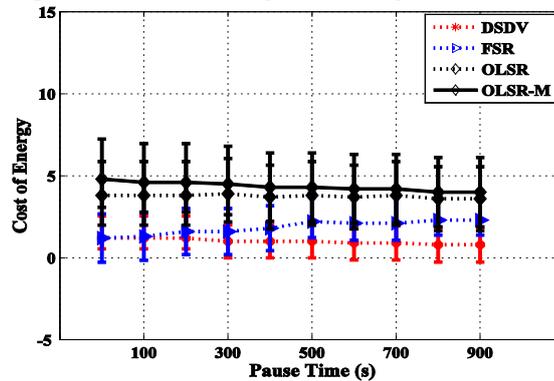

Fig. 13. Time Cost vs Pause Time of Proactive protocols

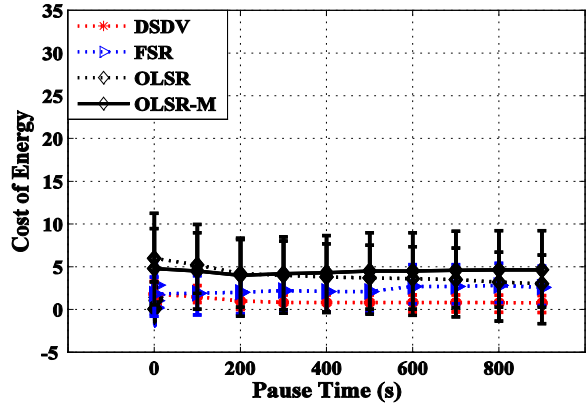
Fig. 14. Time Cost vs Pause Time of Proactive protocols

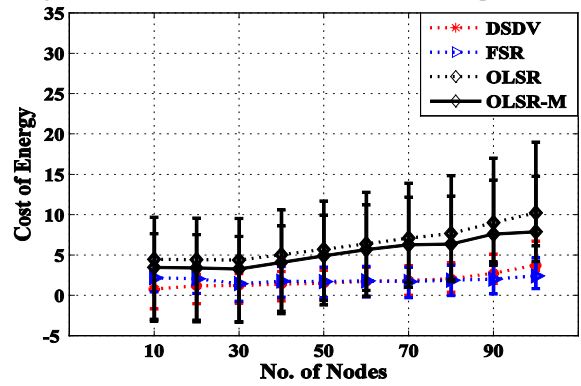
Fig. 15. Time Cost vs Scalability of Proactive protocols

Whereas, FSR reduces congestion by the help of FS technique. The link state packets are not flooded, instead, nodes maintain a link state table based on the up-to date information received from neighboring nodes. Using different exchange periods for different entries in routing table, routing update overhead is reduced. Furthermore, when network size grows large, a GF update plan is used across multiple scopes to keep the overhead low, as portrayed in Fig. 19.

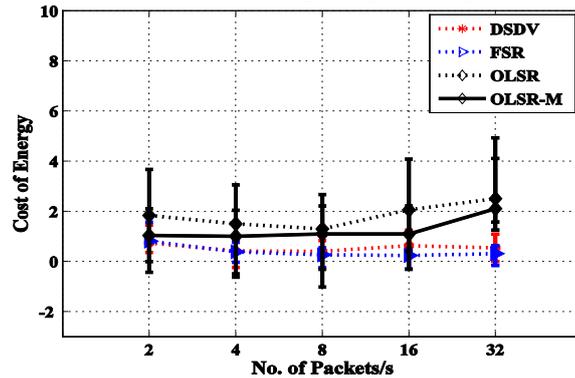
Fig. 16. Time Cost vs Transmission Rate of Proactive protocols

## XV. CONCLUSION AND FUTURE WORK

Energy efficiency and delay reduction are two important factors to check the performance of a protocol in WMhNs. To evaluate these factors, this paper contributes LP_models for WMhNs. To practically examine constraints of respective LP_models over proactive routing protocols, we select DSDV, FSR and OLSR. We relate the effects of routing strategies of respective protocols over WMhNs constraints to check energy efficient and delay reduction of these protocols in different scenarios in NS-2 by considering throughput, E2ED and NRL. DSDV shows more convergence to high dynamicties due to $RU_{tri}$ after detecting link layer feed back and provides optimal solution against constraints of $max\,T_{avg}$. FSR attains highest efficiency in more scalabilies by providing feasible solution against $max\,CE$ constraints due to scope routing. Whereas, OLSR and OLSR-M achieves highest throughput in scalabilies, because of feasible solution through MPRs against all constraints of $max\,CT$.

In future, we are interrested to extend this analysis on the issues addresed in [16-20].